# Correlation of high temperature X-ray photoemission spectral features and conductivity of epitaxially strained $(La_{0.8}Sr_{0.2})_{0.95}Ni_{0.2}Fe_{0.8}O_3/SrTiO_3(110)$

A. Braun[1a], X. Zhang[2,3], Y. Sun[4], U. Müller[5], Z. Liu[6],

S. Erat[1,7], M. Ari[8], H. Grimmer[9], S. S. Mao[2,3], T. Graule[1]

[1] Laboratory for High Performance Ceramics

EMPA – Swiss Federal Laboratories for Materials Testing & Research

CH – 8600 Dübendorf, Switzerland

[2] Department of Mechanical Engineering

University of California Berkeley

Berkeley CA 94720, USA

[3] Environmental Energy Technologies Division

Lawrence Berkeley National Laboratory

Berkeley CA 94720, USA

[4] Stanford Synchrotron Radiation Lightsource

Menlo Park CA 94025, USA

[a] *Corresponding author. Phone +41 (0)44 823 4850, Fax +41 (0)44 823 4150, email: artur.braun@alumni.ethz.ch

[5] Laboratory for Nanoscale Materials Science

Empa – Swiss Federal Laboratories for Materials Testing & Research

CH – 8600 Dübendorf, Switzerland

[6] Advanced Light Source

Ernest Orlando Lawrence Berkeley National Laboratory

Berkeley CA 94720, USA

[7] Department for Nonmetallic Inorganic Materials

ETH Zürich-Swiss Federal Institute of Technology,

CH-8037 Zürich, Switzerland

[8] Department of Physics

Erciyes University

Kayseri 38039, Turkey

[9] Laboratory for Developments and Methods

Paul Scherrer Institut

CH – 5232 Villigen PSI, Switzerland

**Abstract**

Reversible and irreversible discontinuities at around 573 K and 823 K in the electric conductivity of a strained 175 nm thin film of $(La_{0.8}Sr_{0.2})_{0.95}Ni_{0.2}Fe_{0.8}O_{3-\delta}$ grown by pulsed laser deposition on $SrTiO_3$ (110) are reflected by valence band changes as monitored in photoemission and oxygen K-edge x-ray absorption spectra. The irreversible jump at 823 K is attributed to depletion of doped electron holes and reduction of $Fe^{4+}$ to $Fe^{3+}$, as evidenced by oxygen and iron core level soft x-ray spectroscopy, and possibly of a chemical origin, whereas the reversible jump at 573 K possibly originates from structural changes.

Epitaxial strain in films provides opportunities for the miniaturization of systems, devices and components. There is increasing interest for epitaxy of oxide ceramic systems [1]. It has been shown that $SrTiO_3$ under biaxial strain can develop a high temperature ferroelectric phase [2]. Nb doping of $SrTiO_3$ drastically changes the film color, and strained films permit a larger dopant concentration than bulk phases [3]. In-plane compressive strain seems to push away apical oxygen from the $CuO_2$ planes in superconductors, thus enhancing the 2-dimensional character of the dispersion and increasing $T_c$ [4]. The charge ordering temperature in $Pr_{0.5}Ca_{0.5}MnO_3$ was enhanced by epitaxial film growth [5]. Valence band properties of epitaxial LaSrFe-oxide films have been extensively studied with photoemission and oxygen core level spectroscopy [6], but not yet with respect to B-site doping epitaxial strain and high temperature effects. Critical limits are reached in thin films, for instance when temperature activated diffusion processes destroy the discrete structure of multilayer assemblies. Upon annealing, oxygen vacancy formation [7] and ordering [8] must also be taken into account. We present here a thin film study on $(La_{0.8}Sr_{0.2})_{0.95}Ni_{0.2}Fe_{0.8}O_{3-\delta}$ (LSFN), which belongs to the class of multiferroic, strongly correlated electron systems [9]. Our interest for the high temperature properties of nickel substituted LaSrFe-oxides is motivated by their suggested use as cathodes in intermediate temperature solid oxide fuel cells, where addition of Ni causes a substantial increase in conductivity [10]. These materials may have a wider range of applications such as in spintronics [9], and their high temperature (HT) properties could possible play a role such as in *under the hood* applications in automotive technology.

Films with 175 nm thickness and high quality, black luster surface were pulsed laser deposited on single crystal $SrTiO_3$ (110) substrates (a=0.3905 nm) from a $(La_{0.8}Sr_{0.2})_{0.95}Ni_{0.2}Fe_{0.8}O_{3-\delta}$ (LSFN) feed target for 45 minutes at 773 K and 10 mTorr oxygen base pressure, and then in-situ annealed with 250 Torr oxygen. X-ray reflectometry Kiessig oscillations were detected with a high resolution Seifert diffractometer. Structure analysis was carried out with a Philips X'pert diffractometer equipped with a Cu target. No CuK$_{\alpha 2}$ stripping was applied.

Photoemission spectra were collected at BL 8-1 at Stanford Synchrotron Radiation Lightsource in ultrahigh vacuum of 2.1E-09 Torr after careful surface cleaning at 623 K and monitoring volatilization of surface adsorbates with survey scans. After cooling to 323 K, the film was heated to 831 K while spectra were recorded at 60 eV for various T. Arrived at 831 K, the film was flashed to 1023 K for 5 minutes and returned to 828 K. During cooling, spectra were recorded until T = 344 K was reached. The film had then a reddish brownish, translucent color. Best overall coherence was found when for the spectra at 831 K and at 344 K a corrective offset of +0.2 eV and -0.8 eV, respectively, was applied.

Oxygen K-edge and iron L-edge absorption spectra of the as-deposited and annealed film were recorded by target current detection at Beamline 9.3.2, Advanced Light Source, Berkeley, in an UHV chamber with 1.0E-10 Torr base pressure at 300 K. Spectra were collected from 510 eV – 550 eV, 700 eV – 750 eV, and 800 eV – 880 eV in steps of 0.1 eV with energy resolution of around $\Delta E/E \sim 1/10000$.

The electric conductivity as a function of temperature was obtained in the 4-point configuration, with a computer controlled Keithley 2400 sourcemeter and a Keithley 2700 multimeter [11] in a muffle furnace under ambient pressure.

The sharp double peaks (K$_{\alpha 1}$ and K$_{\alpha 2}$) in the x-ray diffractogram (Figure 1) at around 32.45° and 67.8° are identified with the (110) and (220) substrate (S) Bragg reflections, JCPDS 00-035-0734 [12]. The satellites at 31.4° and 31.8° are from the as-deposited (black) and the annealed (red) film, respectively. The inset on the left side magnifies this region of interest, showing that the film grows to a large extent commensurate with the substrate in (110) orientation. The peak at 31.4° is identified with the (110) reflex of the film, and is shifted in the annealed film towards a somewhat larger diffraction angle, suggesting that the film grows with an expanded interlayer distance (1.73 Å), i.e. tensile perpendicular strain and lateral compressive strain, than the corresponding bulk material (1.70 Å), and relaxes to 1.72 Å interlayer distance after annealing towards the equilibrium structure. The x-ray diffractogram of LSFN powder (blue with indexed peaks, Fig. 1) looks similar to that from $SrTiO_3$ nanopowder [12] and is shown for comparison.

Broadening of the film peaks is due to the small film thickness. Kiessig oscillations in the x-ray reflectometry curve account for the quality growth of the as-deposited film and can be modeled with 175 nm thickness (upper right inset).

Not shown here, scanning electron micrographs from the 1300 K annealed film show oblate microcrystals with sizes of around 1-2 micrometer and thickness around 150 nanometer steeping out of the film, suggesting recrystallization during annealing. Within the detection limits, EDAX and XPS did not show any interdiffused Ti from the substrate to the surface of the films, suggesting that the film is thick enough to cover the $SrTiO_3$ substrate, and that the compressive lateral strain in the as-deposited film provides an effective diffusion barrier. At 870 eV, an $L_2$ absorption peak for Ni was found.

Figure 2 displays the valence band photoemission spectra (PES) recorded during heating from 296 K to 831 K, and subsequent cooling to 344 K. The peak assignment is the same as in [6,13], with the $e_g$ band known to arise from doped holes in LSF below -2.0 eV. We identify two sets of spectra that mutually overlap well, this is, spectra for the heating to 296 K < T < 723 K, representative to the fully strained film, and spectra for 831 K > T > 344 K, representative to the partially relaxed film after heating. Between 723 K and 831 K, still in the heating half cycle, there is drastic spectral change, i.e. the lower T spectra have their overall intensity maximum ($t_{2g}$) at -3.8 eV, whereas for T = 831 K, that maximum is at about -5.0 eV.

The $e_g$ band shoulder at -1.9 eV is particularly pronounced for T = 296 K. A second shoulder, left side from the maximum, is observed at about -5.0 eV for 296 < T < 723 K. Here, no shift is observed. This shoulder coincides with the maximum of the spectrum of the T = 558 K maximum ($t_{2g}$).

While the spectra at 831 K and 828 K have not as sharp maxima as those from lower T, the spectra taken at 831 K, which is the last heating temperature and all cooling temperatures look quite similar. Hence, a significant transition has taken place in the film at T > 723 K during heating, but before flashing to 1023 K. These spectra still correspond to the strained film. Note, that after the spectrum for 831 K was recorded, the film was flashed for 5 minutes to 1023 K; this stage represents the relaxed film. Hence, the drastic spectral change between 723 K and 831 K cannot be attributed to the high temperature flashing to 1023 K.

Inspection of all spectra near the Fermi energy $E_F$ from -2.0 eV ≤ E ≤ -1.0 eV shows systematic PES intensity variation with T, which we determined at an arbitrary energy of -1.4 eV and plotted for comparison in Figure 3 as a function of T, reflecting the heating and cooling cycle of the film. The PES intensity has a maximum around 573 K, and then rapidly decreases during further heating at around 723 K – 823 K. Upon cooling, the intensity keeps declining, but passes again an intermediate maximum around 523 K.

The conductivity σ of the film is shown in three subsequent scans for temperatures from 295 K ≤ T ≤ 1298 K, along with the conductivity of a sintered slab from the LSFN powder. The first scan exhibits an overall semiconducting like transport profile. The conductivity at 295 K increases convex to 773 K, and then steeply increases to about 1073 K. After cooling to 300 K, the scan was repeated and a maximum at 1223 K identified. Here, the film temperature did never exceed 1298 K. The second scan follows the global trend of the first scan, while the third scan shows strong deviations. The conductivity increases steeply convex to around 550

K, where it abruptly decreases step-like in a short temperature range. At 598 K, the conductivity then increases concave to about 10.6 kS/m and remains at this value on a plateau. No further increase or decrease is observed up to the terminal temperature of 1298 K. Obviously, after two cycles to high temperatures of 1073 K and then 1298 K, the film does not regain its original maximum conductivity, but remains at a plateau, little less than 1/3 of the overall maximum.

Close inspection of σ around 573 K shows a slight saddle for the first scan, which develops to a clear undulation in the second scan, and to a sharp step in the third scan. In this range, the second scan shows a concave growth, matching that of the third scan very well, with the exception of a constant shift by around 2.0 kS/m up to 823 K. Here, the conductivity of the second scan shows a sharp discontinuity of around Δσ=5.0 kS/m, not observed in the other scans. The first scan, too, shows slight yet noticeable deviation from the typical exponential increase of the conductivity of LSFN at ambient and intermediate T. A relative maximum and minimum around 573 K, as with the second and third scan are not observed, but we notice an inflection point around 598 K, at which the convex conductivity trend switches to an intermediate concave trend until it reaches 623 K. At 673 K, the convex characteristic is again recovered for the first scan conductivity.

The correlation of the peculiarities in the relative PES intensity and the conductivity variation vs. T in Figure 3, is obvious –i) with respect to the initial increase and final decrease of conductivity and PES intensity from ambient temperature to 523 K, -ii) with respect to the intermediate maxima around 573 K, as highlighted with the red circles in Figure 3, and –iii) with respect to the jump feature at ~ 823 K for the second conductivity scan during heating and the PES intensity starting at cooling, as indicated by the vertical red line in Figure 3. The extent of agreement is actually overwhelming, given the different physical mechanisms for PES and conductivity measurements. Similar observations were recently made on a monolithic single crystal [13], underlining that metal oxides with perovskite structure can be meaningful studied ex-situ with PES and correlated with bulk transport properties, including epitaxial strained films.

We believe that the features described in –ii) and –iii) are caused by minute phase transitions in the LSFN, first of which is reversible. Such phase transitions can alter the situation of electron orbitals, and impact the conductivity. While the complexity of phase transitions may increase with decreasing film thickness [15], in monolithic materials they might even not be detected. The first derivative of the dilatometer curve of the sintered LSFN slab (not shown here) exhibits two well resolved intermediate maxima at 540 K and at 870 K, in close proximity to the spectacular changes in conductivity and PES intensity of the films. The conductivity profile of the LSFN sintered bar has no peculiarities, except for an almost unnoticeable step at 486 K with $d\sigma/dT>0$ for T<486, and $d\sigma/dT<0$ for T>486 K. The conductivity profile of the slab is shifted by around 200 K towards lower T, and its conductivity maximum is below that of the strained film.

The change in the conductivity during annealing from the second to third scan, the change in the diffractograms, the PES intensities and the color change shows that the film undergoes severe pathogenesis. The conductivity in LSF is based on electron holes doped by the Sr substitution, which manifests in an extra peak from a spin-up $e_g$ symmetry orbital in the pre-peaks of the oxygen x-ray absorption spectra, in addition to the spin down $e_g$ and $t_{2g}$ bands, which are known for LaFeO3 [16]. While the spectrum of our as-deposited film has a distinct such $e_g$ peak at 525 eV, that peak is absent in the spectrum of the annealed film, Figure 4, revealing that the annealing procedure caused chemical changes in the film that removed the electron holes, most likely by loss of oxygen during annealing. The electron holes in LSF are brought about by $Fe^{4+}$. Annealing causes reduction of the $Fe^{4+}$ towards $Fe^{3+}$ by loss of oxygen. Indeed, the area normalized L-edge spectrum of the annealed film has a relatively larger $L_3$ peak intensity than the as-deposited film, indicating that the Fe is more reduced after annealing, in line with the suggestion that electron holes are removed after annealing, and the conductivity decreases accordingly. We could not make such estimate with respect to the Ni because their $L_3$ edge overlaps with the La $M_{4,5}$ absorption edge.

It remains open to what extent the epitaxial strain in the film has directly changed the film properties with respect to the bulk properties. The temperature range in which σ varies has a direct effect on the transport, potentially caused by lateral compression and vertical expansion of the film lattice. This strain may generally influence the density of states so that the semiconducting type conductivity is suppressed, for example by less overlap of O(2p) orbitals with the Fe and Ni (3d) orbitals, unless thermal activation sets in. Systematic studies on the structure, conductivity and valence bands with x-ray spectroscopy on films with varying substrates and film thickness would likely shed new light on the entanglement of transport, electronic structure and epitaxial strain in films.

## Acknowledgements

Part of this research was carried out at the Stanford Synchrotron Radiation Lightsource, a national user facility operated by Stanford University on behalf of the U.S. Department of Energy, Office of Basic Energy Sciences. Funding by E.U. MIRG # CT-2006-042095, Swiss NSF # 200021-116688, and Erciyes University research fund, contract no. FBA-07-041. The ALS is supported by the Director, Office of Science/BES, of the U.S. DoE, # DE-AC02-05CH11231.

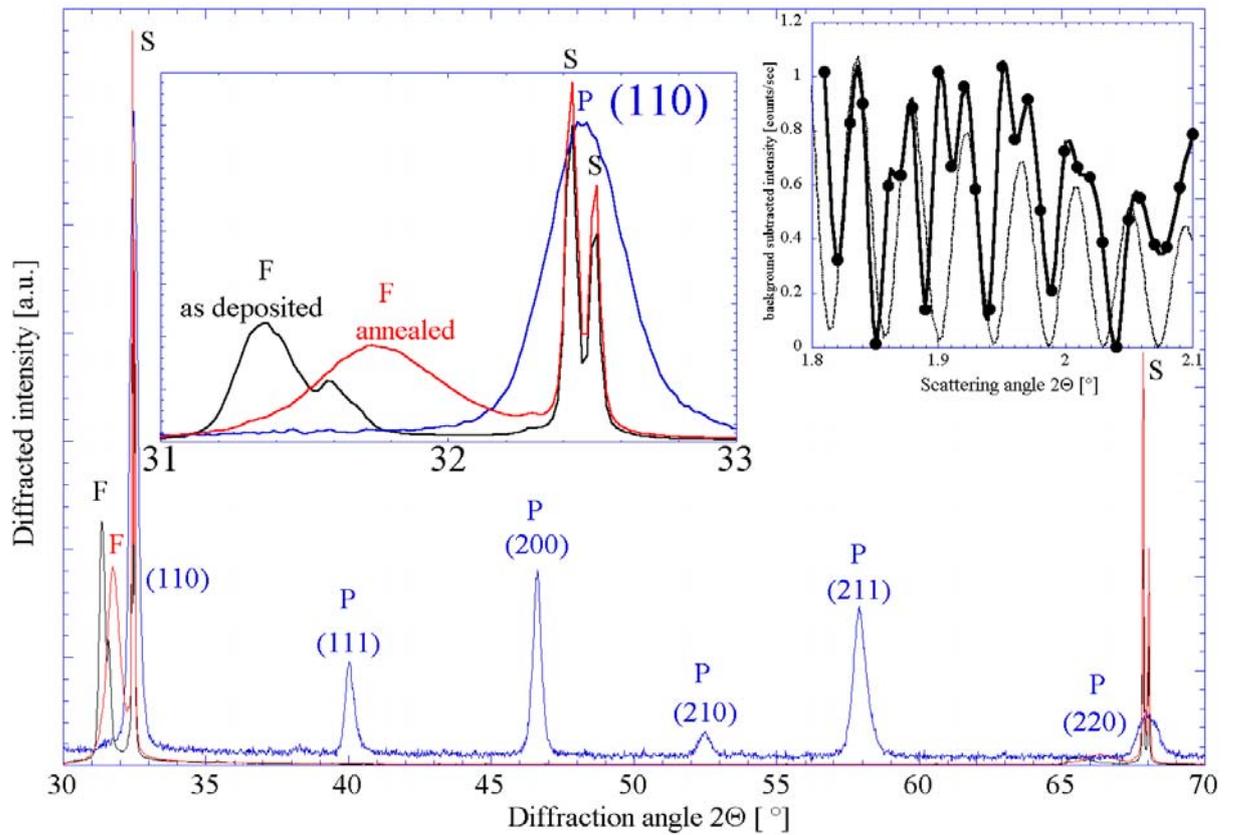

**Figure 1**: Diffractogram of LSFN powder (P, blue with indexed reflections), and as-deposited (F, black) and annealed (F, red) thin films. The left inset shows in magnification the substrate (110) peaks (S) from Cu $K_{\alpha1,2}$ at about 32.45°, and the film (110) reflections between 31° and 32°, respectively. X-ray reflectometry data (right inset) show Kiessig oscillations (filled symbols with cubic spline) which can be modeled with a 175 nm film (thin drawn line).

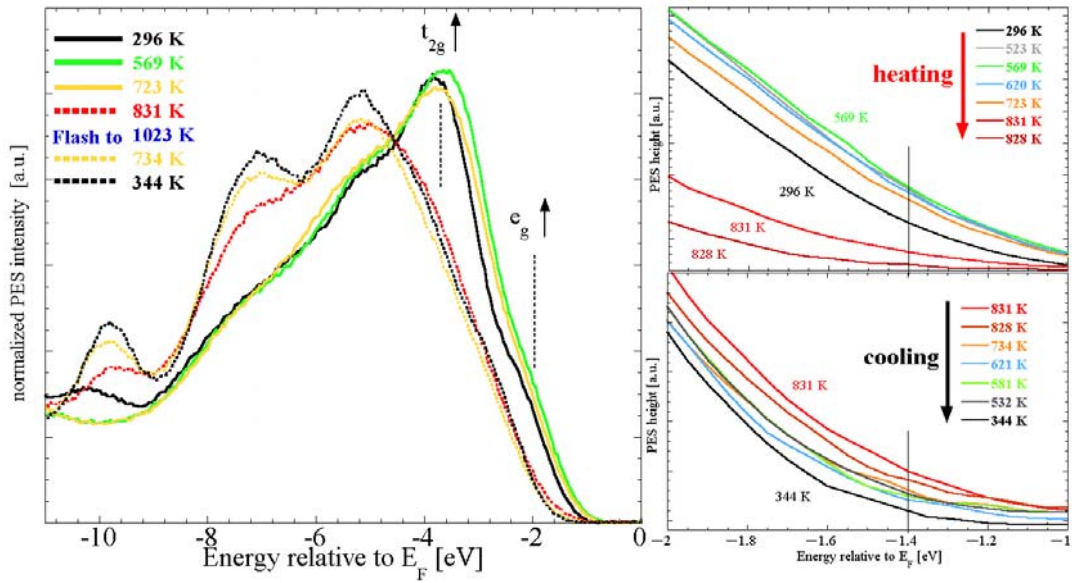

**Figure 2**: Left panel - Normalized and energy corrected photoemission spectra for the heating (left set; 296 K – 723 K) and cooling (right set, 831 K – 344 K). Right panels – Magnified ranges of PES intensity near $E_F$ during heating and cooling for quantitative intensity analysis.

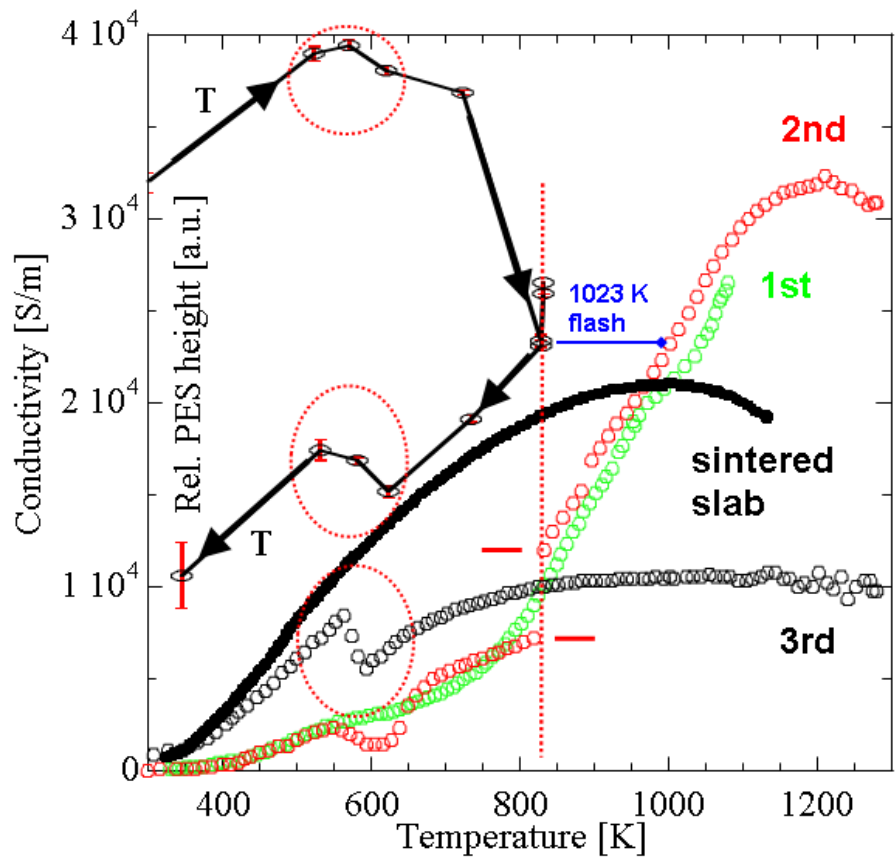

**Figure 3**: Comparison of electric conductivity (lower 3 open symbol curves, solid filled bullet curves) and intensity of the PES spectra (upper left; connected circles with error bars) near the Fermi level (-1.4 eV). The dip in the PES intensity is paralleled by a dip in the conductivity at around 550 K.

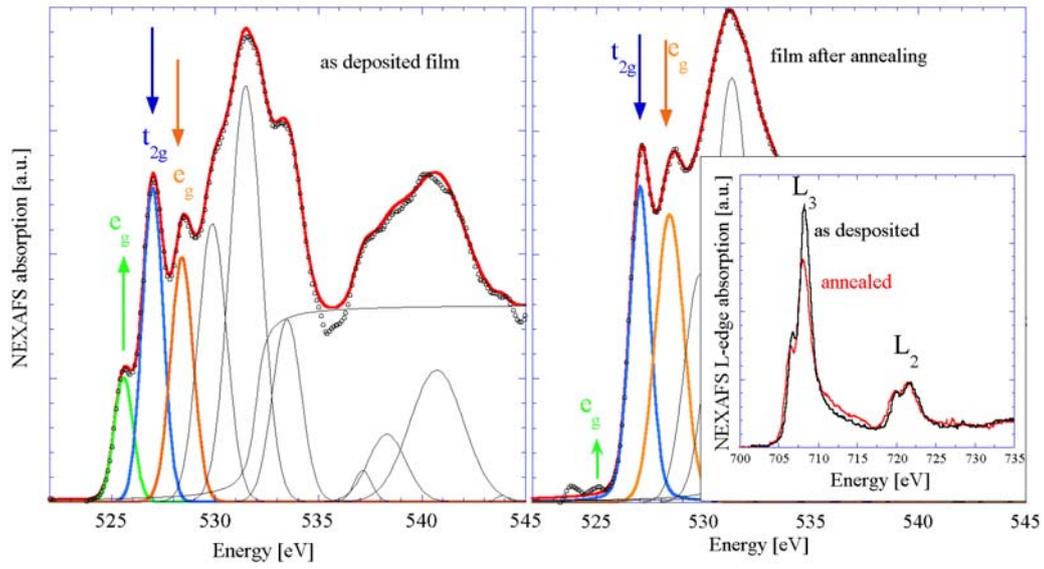

**Figure 4**: Oxygen x-ray absorption spectra of as-deposited (left) and annealed film. Green arrow at 525 eV indicates the position of the realized (upper spectrum) and missing (lower spectrum) hole doped $e_g\uparrow$ states. Inset in right figure shows the Fe L-edges of the films.